\documentclass[pra,a4paper,aps,onecolumn,superscriptaddress,notitlepage,nofootinbib]{revtex4-2}

\usepackage{amsmath,amsthm,amssymb,amsfonts,setspace}
\usepackage{bm}
\usepackage{float}
\usepackage{color}
\usepackage{booktabs}
\usepackage[dvipsnames]{xcolor}
\usepackage[breaklinks=true]{hyperref}
\hypersetup{
  colorlinks   = true, 
  urlcolor     = blue, 
  linkcolor    = blue, 
  citecolor   = red 
}
\usepackage{mathtools}
\usepackage{cleveref} 
\usepackage{braket}
\usepackage{graphicx}
\graphicspath{ {./figs/}} 

\crefformat{equation}{Eq.~(#2#1#3)} 
\crefformat{section}{Section~#2#1#3}
\crefformat{theorem}{Theorem~#2#1#3}
\crefformat{lemma}{Lemma~#2#1#3}
\crefformat{appendix}{Appendix~#2#1#3}
\Crefformat{equation}{Equation~(#2#1#3)}
\crefformat{figure}{Fig.~#2#1#3}
\crefformat{corollary}{Corollary~#2#1#3}
\crefrangeformat{equation}{Eqs.~#3(#1)#4--#5(#2)#6}

\begin{document}

\title{Linear-optical nonlocality without postselection}

\author{Daniel J. Brod}
\affiliation{Instituto de F\'{\i}sica, Universidade Federal Fluminense, Niter\'oi, RJ, 24210-340, Brazil}
\author{Ernesto F. Galv\~ao}
\affiliation{Instituto de F\'{\i}sica, Universidade Federal Fluminense, Niter\'oi, RJ, 24210-340, Brazil}
\affiliation{International Iberian Nanotechnology Laboratory (INL), Av. Mestre Jos\'e Veiga, 4715-330 Braga, Portugal}

\begin{abstract}
Linear optics is one of the major platforms for quantum information processing and experimental tests of quantum foundations. This motivates a thorough exploration of its capabilities and limitations. In this paper we investigate the extent to which passive linear optics, when fed with single-photon Fock states and followed by postselection-free number-resolving measurements, can demonstrate nonlocality. These resources define BosonSampling, a computational model believed to have intermediate computational power, being neither classically simulable nor quantum universal. We prove that, when restricted to only two input photons, this system also has intermediate nonlocality capabilities, achieving a violation of the CHSH inequality no higher than $1+\sqrt{2}$. Although this value has appeared frequently in the literature as achievable, to our knowledge this is the first proof that it is maximal. We show that, as we increase the number of available photons, linear optics can produce violations arbitrarily close to Tsirelson's bound of $2\sqrt{2}$. Finally, we extend the analysis to tripartite nonlocality, leveraging well-known interferometric suppression laws to show how {several} tripartite inequalities can be violated in this setup beyond their bipartite bounds.
\end{abstract}

\maketitle

\section{Introduction} \label{sec:intro}

Bell inequalities find applications both as foundational tests of quantum mechanics \cite{bell1964,Clauser1969} and as operational validations of quantum devices (e.g. for device-independent quantum key distribution \cite{brunner2014}). More generally, demonstrations of nonlocality constitute a natural benchmark of the nonclassical capabilities of specific physical systems, complementing tests of contextuality \cite{budroni_kochen-specker_2022,spekkens_contextuality_2005} and quantum computational advantage \cite{hangleiter_computational_2023,brod2019a}. Here we focus on the nonlocal capabilities of \emph{linear optics}. Photons in passive interferometers are the workhorse of quantum communication and cryptography, as well as the key ingredient for photonic quantum computation. Therefore, it is a natural question how strongly they can provide evidence of the nonlocal nature of quantum mechanics, even in the absence of common nonclassical resources such as entangled states or postselection on outcomes of photon-number measurements.

Consider two parties, Alice and Bob, each of whom can perform one of two measurements ($A_0$ or $A_1$ for Alice, $B_0$ or $B_1$ for Bob). All measurements have outcomes in $\{\pm1\}$. The Bell-CHSH inequality is a bound satisfied by any local realistic theory, given by
$$S_{\textrm{CHSH}} = \langle A_0 B_0\rangle+\langle A_0 B_1\rangle+\langle A_1 B_0\rangle-\langle A_1 B_1\rangle \leq 2.$$
It is well known that quantum mechanics can violate this bound by an appropriate choice of measurements \cite{Clauser1969,bell1964,cirelson1980}, up to a maximum of $2\sqrt{2}$. However, this assumes it is possible to prepare an arbitrary entangled state, as well as perform arbitrary unitary transformations and/or measurements on each individual qubit.

The experimental setup we consider here is typical of linear-optical experiments featuring the usual BosonSampling \cite{Aaronson2013a} constraints: Fock state input, followed by pure passive linear optics, and finally photon-number detection. These resources are not expected to be universal for quantum computation --- universality needs either measurement-induced nonlinearity via feedforward \cite{Knill2001b}, or nonlinear elements \cite{chuang1995}. Therefore, one may reasonably ask whether they are insufficient for other tasks as well, such as demonstrations of nonlocality. Of course, non-universality alone does not settle this question, since maximal violation of CHSH requires only a Bell pair and single-qubit measurements. The
obstruction lies instead in state preparation: passive linear optics cannot
prepare a Bell pair from Fock states deterministically, and heralding one
requires at least four photons \cite{degliniasty2024}.

Nonetheless, it has long been known \cite{popescu1997,yurke1992,yurke1992b} that
nonlocality can be demonstrated in this setting: one can violate CHSH
inequalities with a 2-photon, 4-mode experiment. However, these previous proposals suffer from one of two limitations: either they require the use of postselection, discarding some of the experimental runs, or they do not achieve the maximum possible violation of the CHSH inequality (i.e., Tsirelson's bound \cite{cirelson1980}).

In this work, we investigate to what extent linear optics can demonstrate nonlocality in the absence of postselection, i.e., when retaining the full experimental data. In our view, there are a few reasons to consider setups that use the full data. The first is conceptual: keeping all data avoids having to argue that postselection is benign and does not lead to a detection loophole (though that argument can be made as well, as discussed in
\cite{gebhart2021,gebhart2022,blasiak_safe_2021,blasiak_identical_2025}). Moreover, that
argument relies on global photon-number conservation, which fails as soon as
photons are lost, at which point benign postselection reverts to the usual
detection loophole. A postselection-free violation, in contrast, translates
directly into a detection-efficiency threshold. More importantly, however, we already know that full universal quantum computing is achievable (though inefficiently) by linear optics with postselection \cite{Knill2001b}, and so it is not surprising that postselection enables nonlocality demonstrations with these devices. In contrast, postselection-free linear optics is essentially BosonSampling, which is \emph{not} expected to be universal for quantum computation \cite{Aaronson2013a}. Thus, one might imagine that linear optics is fundamentally incapable  of achieving all values of quantum correlations in the absence of postselection, which arguably would teach us more about the underlying physics (a point discussed recently in \cite{kun2026}, see also \cite{upadhyaya_fundamental_2026} for limits of photonic state preparation). 

In our first result, in \cref{sec:newbound}, we show that, if we prepare a state of two photons, one in each of two modes (together with arbitrarily many extra vacuum modes), and both the preparation of the entangled state and the implementation of measurement settings are performed purely with passive linear optics, then the maximum value for the CHSH correlator is $S_{\textrm{CHSH}}=1+\sqrt{2}$. This value has appeared in the literature several times, for down-conversion photons \cite{popescu1997,zukowski_irrelevance_1999}, independent bosonic sources \cite{yurke1992,yurke1992b,mullin_interference_2008}, pairs of superselection-restricted
states \cite{heaney2010,kumari2024} and single electrons  \cite{dasenbrook2016}, and it has recently been measured \cite{kun2026}. However, in each case it was obtained for a particular scheme or family of measurements, and we are not aware
of any previous proof that it is the maximum achievable value.

Our second result, in \cref{sec:2kphotons}, is to show how one can improve on this value, and in fact get arbitrarily close to Tsirelson's bound, by using an input state with more photons. In particular, we show that if one has $2k$ photons, one can achieve a value of $S_{\textrm{CHSH}}=2\sqrt{2}-\left(\sqrt{2}-1\right)/2^{k-1}$. These results, together, resolve the posed question: even if BosonSampling resources are not universal for quantum computation, they can achieve arbitrarily high violations of the CHSH inequality, though at the cost of using more photons in the state preparation.

Finally, our third result, in \cref{sec:tripartite}, concerns demonstrations of tripartite nonlocality. {We investigate violations of Mermin's \cite{mermin1990a}, Svetlichny's \cite{svetlichny1987} and CWR (Coiteux-Roy, Wolfe and Renou) \cite{coiteux-roy_any_2021,coiteux-roy_no_2021} inequalities using the same resources as described above. The main result here is that, using 6 photons in 6 modes, it is possible to achieve a CWR correlator [i.e., \cref{eq:CWR}] value of 10.24, beating the bound of 10 and thereby certifying genuine tripartite nonlocality}. The key ingredient of this construction is a suppression law of multiport interferometers \cite{tichy2010, crespi2015, dittel_totally_2018}: a symmetry of the input state that forbids certain output patterns. Here, it guarantees that each party can tell from its own photon count whether a run was successful, and that all parties agree. Suppression laws have so far been used mainly to certify photon indistinguishability and to validate boson samplers \cite{tichy_stringent_2014, viggianiello_experimental_2018}; our results give them a different application, as a resource for nonlocality without postselection.

As in the CHSH case, the tripartite results can be improved by using more photons. {Specifically, we show that the CWR correlator approaches 10.65 exponentially fast as the number of photons increases. Interestingly, though, the best violation we found is still below the quantum-mechanical value of $2\sqrt{2}+8\approx 10.83$. A similar result holds for the Svetlichny correlator, though a different scheme does allow the Mermin correlator to achieve its maximal quantum value asymptotically}. To our knowledge, this is the first scheme for demonstration of postselection-free genuine tripartite nonlocality within the experimental setting described here. Although previous work \cite{gebhart2021,gebhart2022} has shown, e.g., how to violate Svetlichny's inequality using postselection, it also relies on additional assumptions (such as no signalling between any two of the three parties), which reduce the scope of the inequality. Heralded GHZ sources based on dedicated auxiliary modes \cite{paesani_scheme_2021, chin_exponentially_2024} could instead be used for event-ready demonstrations of nonlocality; these require auxiliary heralding photons and discard all unheralded runs, and are complementary to the results presented here.

\section{Review: CHSH inequality violation with two photons in four modes} \label{sec:background}

\begin{figure}
    \centering
    \includegraphics[width=0.5\linewidth]{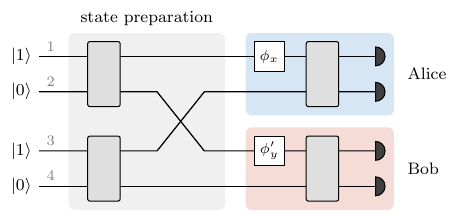}
    \caption{Circuit for violation of the CHSH inequality with two photons, four modes, and pure passive linear optics.}
    \label{fig:chsh}
\end{figure}

We begin by reviewing a well-known construction that has appeared several times in the literature \cite{popescu1997,zukowski_irrelevance_1999,yurke1992,yurke1992b,mullin_interference_2008,kumari2024,dasenbrook2016,heaney2010,kun2026}. The first step is to prepare a state given by
$$\ket{\psi_{in}}=\ket{1010}.$$
This state consists of two photons in four modes, and each label counts the number of photons in that mode. It now propagates through an interferometer which consists of a sequence of a 50:50 beam splitter between modes 1 and 2, another between modes 3 and 4, and a final swap of modes 2 and 3 (cf.\ \cref{fig:chsh}). This prepares the state
\begin{equation} \label{eq:LOBell}
  \ket{\psi_{out}}=\frac{1}{2}\left( \ket{0011} + \ket{0110}+\ket{1100}+\ket{1001}\right)  
\end{equation}
The first two modes of this state are then sent to Alice, and the last two to Bob. Let us reinterpret the above state as a superposition of two entangled states:
$$\ket{\psi_{out}} = \frac{1}{\sqrt{2}}\left( \ket{\mathrm{Bell}} + \ket{\mathrm{fail}} \right).$$
Here, we have
$$\ket{\mathrm{Bell}} = \frac{1}{\sqrt{2}}\left(\ket{0110}+\ket{1001}\right),$$
which can be viewed as a Bell pair in a dual-rail encoding (in which the logical $\ket{0}$ and $\ket{1}$ states correspond to the photonic $\ket{10}$ and $\ket{01}$ states, respectively), and
$$\ket{\mathrm{fail}}=\frac{1}{\sqrt{2}}\left(\ket{0011}+\ket{1100}\right).$$

Let us ignore the $\ket{\mathrm{fail}}$ component for a moment. If Alice and Bob shared the $\ket{\mathrm{Bell}}$ state, they could easily violate CHSH maximally by performing the standard CHSH measurements in the dual-rail encoding, which can be achieved by a sequence of beam splitters, phase shifters and number-resolving measurements, as represented in \cref{fig:chsh} \cite{chuang1995}.

There are now two strategies to deal with the $\ket{\mathrm{fail}}$ component:
\begin{itemize}
    \item[(i)] \textbf{Postselection}. Whenever Alice and Bob detect either 0 or 2 photons, they discard the data from that run. Usually, one avoids relying on postselection for Bell inequality violations, since it can open a detection loophole \cite{pearle1970,garg1987,popescu1995}. However, in this case, postselection is actually benign \cite{blasiak_safe_2021,gebhart2022}. The reason is that, due to global particle-number conservation, the decision to discard a run is perfectly correlated between Alice and Bob --- whenever Alice receives 0 photons, Bob receives 2, and vice versa, and this acts as a failure flag that they both share.
    \item[(ii)] \textbf{No postselection}. Alternatively, Alice and Bob can just incorporate every outcome in their data and accept that some of their outcomes will be useless.
\end{itemize}

By using strategy (ii), Alice and Bob achieve a CHSH value of $S_{\textrm{CHSH}} = 1+\sqrt{2}\approx 2.4142$. Intuitively, this follows from the fact that they get a successful run 50\% of the time, and in those cases they violate CHSH maximally, while in the other 50\% they know that the run has failed, so they can choose to output answers for which $S_{\textrm{CHSH}}=2$. This value has surfaced repeatedly in the long debate on the nonlocality of a single photon \cite{tan1991,hardy1994,peres1995,vaidman1995,greenberger1995} and, most recently, in an experiment reported in \cite{kun2026}, where $S_{\textrm{CHSH}}=2.23\pm0.07$ was measured without postselection (versus $2.71\pm0.09$ with it).

\section{Bound for CHSH violation with two photons} \label{sec:newbound}

In the previous section we described a method to obtain a CHSH value of  $S_{\textrm{CHSH}}=1+\sqrt{2}$ with postselection-free linear optics. Is this the optimal violation, when allowing for arbitrary choices of interferometers both for state preparation and for the measurement settings by Alice and Bob? We now prove that this is the case, as long as there are no more than two photons during the state preparation phase, and that Alice and Bob are not allowed to use coherent superpositions of different photon numbers (such as coherent states) during their respective measurement phases. The intuition behind the result is that there is a trade-off: Alice and Bob cannot arbitrarily reduce the probability of failed outcomes (i.e., when both photons go to a single party) while also keeping the shared state highly entangled when the runs do succeed.

Consider an initial state composed of two photons, one in each mode; let us write it as $a^\dagger_1 a^\dagger_2\ket{0}$. We assume that we have access to extra modes initialized in the vacuum state, for an (arbitrary) total of $M$ modes. A general $M$-mode interferometer $U$ is then applied to the state, and subsequently some modes, say $\{1,\ldots,m\}$, are sent to Alice, and the remainder to Bob. The state, at this stage, can be written as
$$\left(\sum_{i=1}^{m} U_{i,1} a^\dagger_i+\sum_{i=m+1}^{M} U_{i,1}a^\dagger_i  \right)\left(\sum_{j=1}^{m} U_{j,2} a^\dagger_j+\sum_{j=m+1}^{M} U_{j,2}a^\dagger_j  \right) \ket{0}.$$
This can be written as 
$$\ket{\psi_{20}}+\ket{\psi_{02}}+\ket{\psi_{11}},$$
where
\begin{align*}
\ket{\psi_{20}} & =\sum_{i=1}^{m} \sum_{j=1}^{m} U_{i,1} U_{j,2} a^\dagger_i a^\dagger_j\ket{0},\\
\ket{\psi_{02}}&=\sum_{i=m+1}^{M} \sum_{j=m+1}^{M} U_{i,1} U_{j,2} a^\dagger_i a^\dagger_j\ket{0},
\end{align*}
correspond to the sectors where either Alice or Bob receives both photons, with respective probabilities $p_{20}=\|\ket{\psi_{20}}\|^2$ and $p_{02}=\|\ket{\psi_{02}}\|^2$, and
\begin{equation*}
    \ket{\psi_{11}} =\sum_{i=1}^{m} \sum_{j=m+1}^{M} (U_{i,1} U_{j,2}+U_{i,2} U_{j,1}) a^\dagger_i a^\dagger_j\ket{0},
\end{equation*}
corresponds to the sector where they both received one photon each, which occurs with probability $p_{11}=\|\ket{\psi_{11}}\|^2$. Naturally, unitarity of $U$ implies $p_{02}+p_{20}+p_{11}=1$.

No matter how Alice and Bob encode their measurement settings in their interferometers, by assumption the photon number is conserved locally (i.e., on each side). This means that there is no interference between sectors with different local photon numbers, and that propagates into the final expectation of the CHSH observable, which decomposes as
$$S_{\textrm{CHSH}} = p_{02} S_{02} + p_{20} S_{20} + p_{11} S_{11},$$
where $S_{ij}$ is the CHSH value obtained by Alice and Bob in the $\{i,j\}$ sector. 
When Bob receives 0 photons, he cannot actually perform any nontrivial measurements, since the only possible outcome is vacuum, and similarly when Alice receives 0 photons. This means that we can set $S_{02}=S_{20}=2$---since in those cases there will be no correlations at all, they might as well output outcomes that achieve the classical bound. Thus we can write
$$S_{\textrm{CHSH}} = 2(p_{20}+p_{02}) + p_{11} S_{11}.$$

Our goal now is to show that there is a trade-off between $p_{11}$ and $S_{11}$. To see that, write 
\begin{equation*}
    \ket{\psi_{11}} = \left(\sum_{i=1}^{m} U_{i,1}  a^\dagger_i \right)\left(\sum_{j=m+1}^{M} U_{j,2}  a^\dagger_j \right) \ket{0} + \left(\sum_{i=1}^{m} U_{i,2}  a^\dagger_i\right)\left(\sum_{j=m+1}^{M} U_{j,1}  a^\dagger_j \right) \ket{0}.
\end{equation*}
In other words: the state is a sum of two (subnormalized) tensor-product states. Therefore, we can write it in its Schmidt form, with Schmidt rank at most 2:
\begin{equation*}
    \ket{\psi_{11}} = \alpha \ket{e_1, f_1}+\beta \ket{e_2,f_2},
\end{equation*}
where $\left\{\ket{e_1},\ket{e_2}\right\}$ are two orthogonal vectors on Alice's side, and similarly for $\left\{\ket{f_1},\ket{f_2}\right\}$ on Bob's side, and where $p_{11} = \left|\alpha\right|^2+\left|\beta\right|^2$.
Using a result due to Gisin \cite{gisin_bells_1991} (see also \cite{horodecki_violating_1995}), it follows that
$$p_{11} S_{11} \leq 2\sqrt{p_{11}^2+4|\alpha \beta|^2}.$$

The final step is now to bound the product $|\alpha \beta|$. To that end, define the (subnormalized) single-photon states
\begin{equation*}
    \ket{u_k} = \sum_{i=1}^{m} U_{i,k}\, a^\dagger_i \ket{0}, \qquad
    \ket{v_k} = \sum_{j=m+1}^{M} U_{j,k}\, a^\dagger_j \ket{0}, \qquad k = 1,2,
\end{equation*}
so that
\begin{equation*}
    \ket{\psi_{11}} = \ket{u_1}\ket{v_2} + \ket{u_2}\ket{v_1}.
\end{equation*}
Let $t_k = \|u_k\|^2$ denote the probability that photon input in mode $k$ is routed to Alice. The unitarity of $U$ implies $\|v_k\|^2 = 1 - t_k$.

For a state of this form, the product of the Schmidt coefficients can be bounded (via the concurrence) as \cite{fu2001,hill1997}
\begin{align*}
    |\alpha\beta|^2
    &= \left( \|u_1\|^2 \|u_2\|^2 - |\braket{u_1|u_2}|^2 \right)
       \left( \|v_1\|^2 \|v_2\|^2 - |\braket{v_1|v_2}|^2 \right) \\
    &\leq t_1 (1 - t_1)\, t_2 (1 - t_2) \\
    &\leq \frac{1}{16}.
\end{align*}

If $p_{11} \leq \frac{1}{2}$, Tsirelson's bound $S_{11} \leq 2\sqrt{2}$ gives
\begin{equation*}
    S_{\mathrm{CHSH}} \leq 2(1 - p_{11}) + 2\sqrt{2}\, p_{11} \leq 1 + \sqrt{2}.
\end{equation*}
If instead $p_{11} \geq \frac{1}{2}$, then
\begin{equation*}
    S_{\textrm{CHSH}} \leq 2(1 - p_{11}) + 2\sqrt{p_{11}^2 + \tfrac{1}{4}},
\end{equation*}
which has a maximum $1 + \sqrt{2}$ at $p_{11} = \frac{1}{2}$, which concludes our proof.

\section{CHSH inequality violation with $n>2$ photons} \label{sec:2kphotons}

We now show that it is possible to obtain values of $S_{\textrm{CHSH}}$ arbitrarily close to Tsirelson's bound by using more than two photons and four modes. First, suppose that we have an \emph{even} total photon number. In this case, we can bootstrap strategy (ii) described in \cref{sec:background}, as follows. 

\begin{figure}[b]
    \centering
    \includegraphics[width=0.6\linewidth]{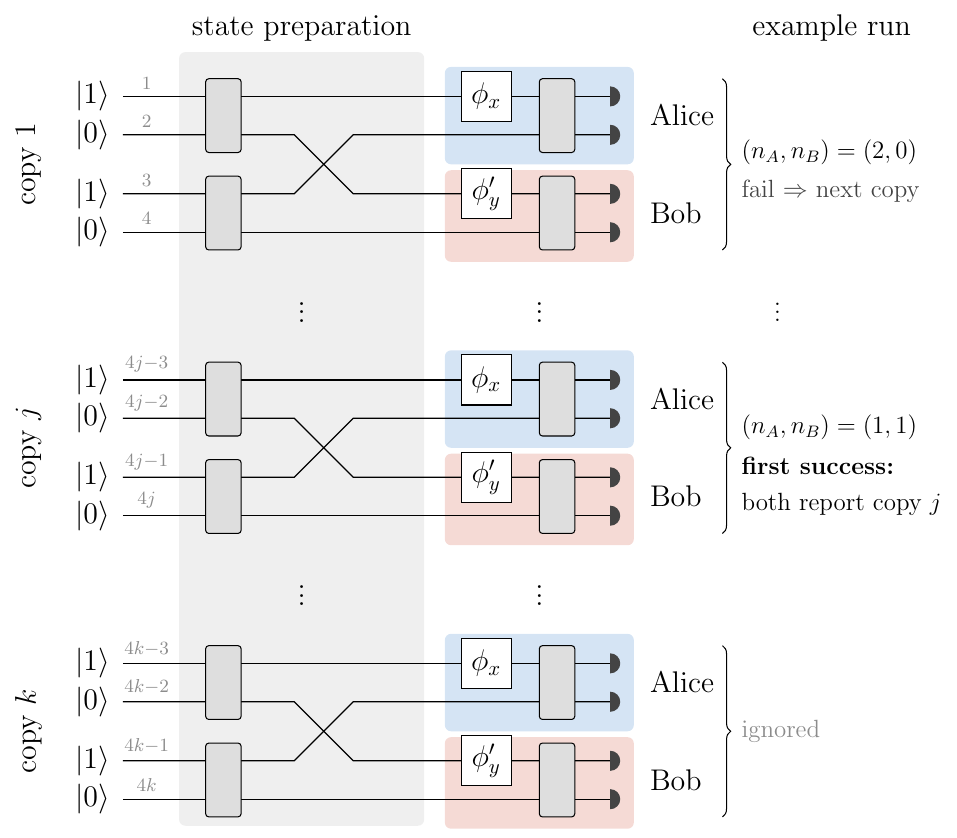}
    \caption{Ladder construction for violation of the CHSH inequality with $2k$ photons.}
    \label{fig:ladder}
\end{figure}

We begin by performing \emph{a ladder} of $k$ parallel instances of the interferometer described in \cref{fig:chsh}, resulting in the preparation of the state
\begin{equation} \label{eq:kBell}
    \frac{1}{\sqrt{2^k}} (\ket{\mathrm{Bell}}+\ket{\mathrm{fail}})^{\otimes k},
\end{equation}
which uses $2k$ photons and $4k$ modes. Alice then receives the first two modes from each state in the tensor product above, and the remaining $2k$ modes go to Bob. They perform exactly $k$ parallel copies of the protocol described in \cref{sec:background}, measuring the same observable in all $k$ runs, according to some canonical ordering established upfront. Finally, they output the value of the \emph{first} successful measurement, i.e., where each detected a single photon. Due to number conservation within each term of the tensor product in \cref{eq:kBell}, all failures and successes are perfectly correlated, and the ordering ensures that they will output the measurement outcomes coming from the same Bell pair. If every single event fails, they both just output the value $+1$. The full scheme is represented in \cref{fig:ladder}.

By construction, Alice and Bob are guaranteed to report measurements performed on a valid dual-rail Bell pair, except in the case where all runs fail, which occurs with probability $1/2^k$. It is easy to work out that the overall value of the CHSH correlator of this concatenated postselection-free strategy is
$$S_{\textrm{CHSH}}  = 2\sqrt{2} \left(1-\frac{1}{2^k}\right)+2 \frac{1}{2^k},$$
which converges to $2\sqrt{2}$ exponentially fast with the number of repetitions\footnote{Incidentally, the rate $1-2^{-k}$ is the same geometric ladder that
governs boosted Bell-state measurement \cite{grice_arbitrarily_2011,ewert_3_2014},
which has been demonstrated beyond 50\% success \cite{bayerbach_bell-state_2023}.}. For $k=2$ (4 photons, 8 modes) this gives $S_{\textrm{CHSH}}=(1+3\sqrt2)/2\approx 2.6213$.

These results contrast with previous postselection-free many-particle schemes, whose
violations decrease as particles are added \cite{mullin_interference_2008,heaney2010}, which led to the suggestion that the superselection rule prevents a maximal violation for any
particle number \cite{heaney2010}. Our construction shows that this decrease in violation with photon number is specific to
those schemes. Note, however, that ours is not a heralded source of Bell pairs:
the success flags are only read at measurement time, so the successful pairs
cannot be used in subsequent operations.

\begin{figure}
    \centering
\includegraphics[width=0.6\linewidth]{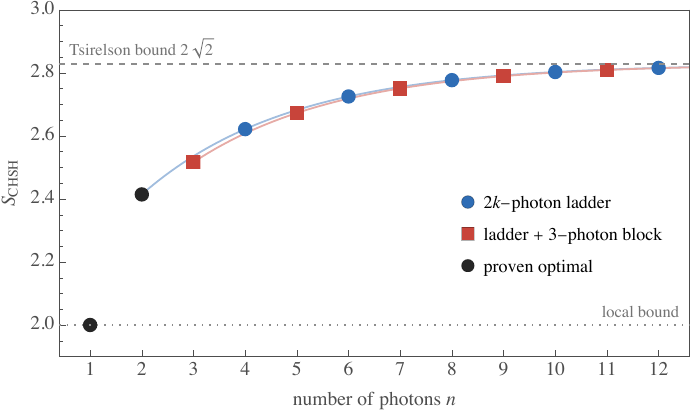}
    \caption{Postselection-free violation of the CHSH inequality as a function of photon number $n$. Curves follow the points as continuous functions of $n$.}
    \label{fig:chshvsn}
\end{figure}

If the total photon number is odd, one approach is to ignore one photon and perform the same strategy as above. However, it is possible to use a different approach and do a little better. As we describe in detail in \cref{apx:odd}, we have obtained a construction that uses 3 photons in 4 modes and achieves a value of $S_{\textrm{CHSH}} \approx {2.52}$. For an arbitrary odd photon number $n\geq 5$, we can then mix and match the two constructions: $(n-3)/2$ parallel repetitions of the construction of \cref{fig:chsh}, and for the final 3 photons we perform the construction from \cref{apx:odd}. In \cref{fig:chshvsn} we plot the best value of $S_{\textrm{CHSH}}$,  as a function of the number of photons, by the methods described here.

\section{Tripartite nonlocality} \label{sec:tripartite}

We now move to the case of tripartite nonlocality, in the form of {two well-known inequalities, Mermin's \cite{mermin1990a} and Svetlichny's \cite{svetlichny1987}, as well as a more recent one due to Coiteux-Roy, Wolfe and Renou (CWR) \cite{coiteux-roy_any_2021,coiteux-roy_no_2021}}.

Mermin's inequality was proposed in \cite{mermin1990a}. It considers three parties, Alice, Bob and Charlie, each of whom can make one of two measurements with outcomes $\pm 1$. The Mermin correlator is given by
\begin{equation}\label{eq:Merm}
S_{\textrm{Merm}} = \left \langle A_0 B_0 C_0 \right \rangle - \left \langle A_0 B_1 C_1 \right \rangle - \left \langle A_1 B_1 C_0 \right \rangle - \left \langle A_1 B_0 C_1 \right \rangle.
\end{equation}
The upper bound achieved for $S_\textrm{Merm}$ by local realistic theories is 2. This can be violated up to $2 \sqrt{2}$ if two of the three parties share a Bell pair and perform a usual CHSH strategy (while the third party always outputs the value +1, say). This can then be violated further by using a genuine tripartite entangled state. In particular, using a GHZ state one can achieve the algebraic maximum of 4 \cite{mermin1990a}.

Svetlichny's inequality was proposed in \cite{svetlichny1987}, as a more stringent demonstration of genuine tripartite nonlocality. The Svetlichny correlator is given by
\begin{equation} \label{eq:svet}
    S_{\textrm{Svet}}=\left \langle A_0 B_0 C_0 \right \rangle + \left \langle A_0 B_0 C_1 \right \rangle + \left \langle A_1 B_0 C_0 \right \rangle + \left \langle A_0 B_1 C_0 \right \rangle-\left \langle A_1 B_0 C_1 \right \rangle - \left \langle A_0 B_1 C_1 \right \rangle - \left \langle A_1 B_1 C_0 \right \rangle - \left \langle A_1 B_1 C_1 \right \rangle.
\end{equation}
The maximum achievable value for $S_{\textrm{Svet}}$ is 4 both for a local realistic theory and for any biseparable strategy, i.e., one where, in each run, a single pair of parties (possibly chosen at random) may share arbitrary, even signalling, correlations, while the third party is only classically correlated with them. By using a GHZ state and suitable  measurements in the equator of the Bloch sphere, one can achieve $S_{\textrm{Svet}} = 4\sqrt{2}$.

{Both Mermin's and Svetlichny's inequalities were used in the past as witnesses of tripartite nonlocality, but this interpretation is subject to some important caveats. Mermin's inequality, for instance, can be violated partially if two of the three parties share a Bell pair, and up to the algebraic maximum if two parties share PR-box type correlations \cite{popescu1994}. Similarly, correlations obtained from Bell pairs distributed between the parties can be genuinely tripartite nonlocal in Svetlichny's sense~\cite{contreras-tejada2021}, and three PR-boxes, one per pair, reach the algebraic maximum of 8 of Svetlichny's inequality, as neither scenario falls under the definition of biseparable. These limitations of the previous inequalities motivated the work of Coiteux-Roy, Wolfe and Renou \cite{coiteux-roy_any_2021,coiteux-roy_no_2021}, who obtained a set of inequalities that cannot be violated by any strategy where Alice, Bob and Charlie share arbitrary bipartite resources of any causal theory (classical, quantum, or post-quantum such as PR-boxes), supplemented by global shared randomness. Thus, here we view the CWR inequality as a more stringent witness of genuine tripartite nonlocality\footnote{Strictly speaking, the CWR and Svetlichny inequalities are incomparable. Correlations from distributed PR-boxes can violate Svetlichny's, which CWR was designed to rule out. Conversely, CWR is formulated within Local Operations and Shared Randomness (LOSR), so its inequality can be violated if two of the parties may communicate freely, whereas Svetlichny's cannot.}, though violations of both Mermin and Svetlichny above their bipartite bounds can still be used as device-independent witnesses of genuine tripartite entanglement if one assumes quantum theory (see the discussion in \cite{coiteux-roy_any_2021,coiteux-roy_no_2021} and references therein).}

{The CWR inequality requires a slightly different setup, where one party (say, Bob) is allowed to measure according to three distinct settings, labelled $\{0,1,2\}$. Then we write:}
\begin{align}
    S_{\textrm{CHSH}}^{C_1=1} &=
     \left \langle A_0 B_0 \right \rangle_{C_1=1} + \left \langle A_0 B_1 \right \rangle_{C_1=1} + \left \langle A_1 B_0 \right \rangle_{C_1=1} - \left \langle A_1 B_1 \right \rangle_{C_1=1}, \\
       S_{\textrm{same}} &=     \left \langle A_0 B_2 \right \rangle + \left \langle B_2 C_0 \right \rangle.
\end{align}
{We can understand $S_{\textrm{CHSH}}^{C_1=1}$ as the standard CHSH correlator between Alice and Bob whenever Charlie makes measurement 1 and obtains outcome +1, whereas $S_{\textrm{same}}$ just measures the correlation between the three outcomes when Bob makes measurement 2,  obtaining a maximum value whenever the three outputs agree. Proposition 2 in \cite{coiteux-roy_no_2021} then shows that, under the assumption that $\left \langle C_1 \right \rangle = 0$\footnote{This condition might be a problem in an experimental setting, since one would never observe that exactly. However, \cite{coiteux-roy_any_2021} has a more complete analysis that drops this requirement, though we do not pursue that case here.} and that Alice, Bob and Charlie do not share any tripartite nonclassical resource,
\begin{align} \label{eq:CWR}
   S_{\textrm{CWR}}= S_{\textrm{CHSH}}^{C_1=1} + 4 S_{\textrm{same}} \leq 10.
\end{align}
They also show that, using a GHZ state, quantum mechanics can achieve a value of $2\sqrt{2}+8\approx 10.83$, and the algebraic maximum is 12.}

Here, we show how {all three inequalities} can be violated beyond their bipartite bounds (i.e., $2\sqrt{2}$ for Mermin, $4$ for Svetlichny {and 10 for CWR}) using linear optics with Fock state inputs and measurements.

The first approach could be to generalize the strategies used for violation of CHSH in \cref{sec:background}, which would suggest using $n=3$ photons in $m=6$ modes \cite{yurke1992,yurke1992b,kumar_experimental_2023,piccolini_colloquium_2026}. One might then put each photon in superposition of two modes, and give two out of the six modes to each party. This does produce a state similar to \cref{eq:LOBell}, which consists partly of a dual-rail-encoded GHZ state and partly of a failed state:
\begin{align}
  \ket{\psi}
  &= \frac{1}{2\sqrt2}\left(\ket{101010} + \ket{010101}\right)+\frac{1}{2\sqrt2}\left(\ket{111000}+\ket{110001}+\ket{011100}+\ket{001110}+\ket{100011}+\ket{000111}\right).
\end{align}
However, this approach does not straightforwardly lead to a violation. Strategy (ii), of incorporating all data into the expectation value and hoping that the failed outcomes do not drown out the signal from the successful sector, numerically does not yield any violation at all. Strategy (i), of using benign postselection, also cannot be immediately extended, because in every failed outcome only two of the three parties are aware that that run should be discarded. The latter argument is not a proof that there is no benign postselection---indeed, recently, it was shown that it is possible to obtain a violation using postselection as long as more than half of the parties can tell that the run was a failure \cite{gebhart2021,gebhart2022}. Nonetheless, as we argued previously, we see a few drawbacks of following this route: first, the multipartite postselection argument of \cite{gebhart2021,gebhart2022} is arguably more complex than the bipartite one in \cref{sec:background}, as well as the ones we will describe shortly. Second, their argument also requires no-signalling between any two of the three parties, which reduces the scope of Svetlichny's inequality in particular. Finally, the fact that linear optics can perform more tasks when augmented with postselection is perhaps not so surprising due to the universality results of \cite{Knill2001b}, and we wish to investigate whether it retains these capabilities in the absence of postselection.

\begin{figure}
    \centering
    \includegraphics[width=0.5\linewidth]{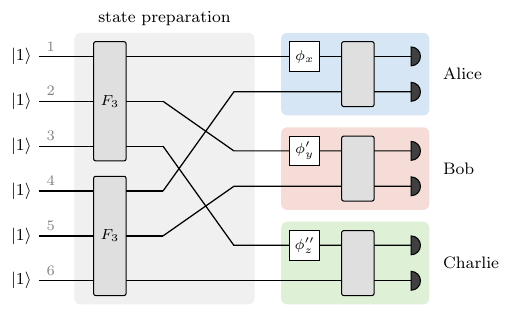}
    \caption{Linear-optical circuit for violation of tripartite inequalities.}
    \label{fig:tripartite}
\end{figure}

Consider then the circuit shown in \cref{fig:tripartite}. This circuit initializes 6 photons, one in each mode. Three of these modes get sent into a tritter, i.e., a 3-mode quantum Fourier transform (QFT), while the other three get sent into a second, identical tritter. Finally, one output mode of each tritter is sent to each of the parties. The final shared state can be written as follows:
$$-\frac{\sqrt{2}}{3\sqrt{3}}\left(\ket{310101}+\ket{131010}+\textrm{perm.}\right)+\frac{1}{3}\ket{111111}+\frac{2}{9}\left(\ket{330000} + \ket{300300} + \textrm{perm.}\right),$$
where the permutations indicated correspond to permutations of the states between the three parties. Note that, due to a well-known suppression law of the tritter \cite{tichy2010}, no states in the output contain exactly two photons in the same mode (a similar zero-transmission law for Fourier interferometers underlies heralded GHZ generation schemes \cite{paesani_scheme_2021}; there, success is flagged by dedicated heralding modes, whereas here it is flagged locally by each party’s own photon count). We now claim that this state can be reorganized as
$$-\frac{2}{3\sqrt{3}} \left(\ket{\mathrm{GHZ}_1}+\ket{\mathrm{GHZ}_2}+\ket{\mathrm{GHZ}_3}\right)+\frac{\sqrt{5}}{3}\ket{\mathrm{fail}}$$
where $\ket{\mathrm{GHZ}_i}$ are three copies of GHZ states using different encodings, which we will discuss shortly, and $\ket{\mathrm{fail}}$ is the state that includes all failure modes. 

Let us first focus on the failure modes. Note that, in every state in $\ket{\mathrm{fail}}$, each party gets 0, 2, 3 or 6 photons, whereas the GHZ-like terms all correspond to each party getting exactly 1 or 4 photons. This is the failure flag for this strategy: whenever each party receives a number of photons different from 1 or 4, they know (as do the others) that the run has failed. As the state above shows, success is perfectly correlated over all parties, so we might here consider a similar postselection strategy as described in \cref{sec:background}. However, as justified in \cref{sec:background}, instead of postselection we focus on violations even when all data are kept. Here, this can be achieved by the three parties agreeing on deterministic outputs that saturate {the local bounds of the inequalities} whenever they observe a failure. As long as they can violate the inequalities for the successful events, this will convert into a smaller violation when all data are kept, as in the bipartite case of \cref{sec:background}.

Now we consider the $\ket{\mathrm{GHZ}_i}$ events. Consider first the state $\ket{\mathrm{GHZ}_1}=\frac{1}{\sqrt{2}} \left(\ket{310101}+\ket{131010}\right)$. This is a GHZ state if we consider the last two qubits to be in a standard dual-rail encoding, whereas the first qubit is encoded in the 4-photon basis $\{\ket{31},\ket{13}\}$. The other two states $\ket{\mathrm{GHZ}_2}$ and $\ket{\mathrm{GHZ}_3}$ are the analogous cases where the second and third qubit are the ones encoded in the 4-photon basis, respectively. 

{For Mermin and Svetlichny,} the crucial observation that makes this scheme work is that the measurement-setting interferometers, shown in the colored boxes of \cref{fig:tripartite}, which consist of phase shifters followed by balanced beam splitters and number-resolving detectors, act as measurements in the equator of the Bloch sphere on both qubit encodings. In the $\{01,10\}$ encoding, it is a measurement that makes an angle $-\phi$ with the $X$ direction, where $\phi$ is the angle of the phase shifter. In the $\{31,13\}$ encoding, after the beam splitter we can observe four possible events, corresponding to $13,31,04$ and $40$ detections (the $22$ outcome is suppressed). By coarse-graining the pair $40, 04$ as the qubit outcome $0$ and the remainder as the qubit outcome $1$, we again get a measurement in the equator of the Bloch sphere. However, because the phase imparted by the phase shifter is proportional to the photon number in the mode, we get an effective measurement that makes an angle of $2\phi$ with the $X$ direction instead.

In principle, with arbitrary measurements in the equator of the Bloch sphere, we can violate both Mermin's and Svetlichny's inequalities maximally. However, here we find an obstruction to achieving the upper bound in our scheme: the three parties do not know in which of the three $\ket{\mathrm{GHZ}_i}$ events they are until after the measurements, and therefore they must use the same phase shifter angle regardless of which encoding their qubit will arrive in---but the same phase shifter implements different measurements in the two encodings. If we now maximize over the choices of angles (reported in \cref{apx:tripartite}), as well as the choice of assignments between measurement outcomes and final $\pm 1$ output values, and finally accounting for the probability of the failed events, we arrive at the postselection-free values
\begin{align}
    S_{\textrm{Merm}} & = \frac{4}{9} \times 3.5+\frac{5}{9} \times 2 \approx 2.67,\\
    S_{\textrm{Svet}} & = \frac{4}{9}\times 5 + \frac{5}{9} \times 4 \approx 4.44.
\end{align}
Note that $S_{\textrm{Svet}}$ is a violation of the bound of 4. Numerical evidence suggests that these values cannot be improved further by changing the beam splitter reflectivities.

{For CWR, the measurement strategy is a bit more complex, since measurements in the equator of the Bloch sphere no longer suffice, and Bob needs three distinct settings. This in turn requires tuning the parameters of both the beam splitter and the phase shifter, which corresponds to implementing arbitrary single-qubit measurements in the $\{01,10\}$ encoding, though it causes states in the $\{31,13\}$ encoding to leak out of it. It is also important for Charlie to output a random bit whenever he observes a failure on setting 1, to ensure that $\langle C_1 \rangle=0$. In \cref{apx:tripartite} we report a valid set of parameters which leads to}
\begin{equation}
    S_{\textrm{CWR}} = \frac{4}{9} \times 10.54 + \frac{5}{9} \times 10 \approx 10.24.
\end{equation}

We remark that heralded GHZ sources \cite{paesani_scheme_2021,chin_exponentially_2024} allow event-ready Bell tests with Fock inputs and linear optics. The scheme of \cite{chin_exponentially_2024} uses the same six photons as ours, with one H and one V photon per input port, but spends three of them as heralds, succeeds with probability 1/32 \cite{chin_exponentially_2024,varnava2008}, includes feed-forward corrections, and requires discarding unheralded runs. Here, no photon is used as a herald and no run is discarded, at the cost of the angle constraint discussed above, and there is no need for feed-forward corrections (heralded schemes without feed-forward also exist, though at lower success probability \cite{gubarev_improved_2020}).

Since there is a clear failure/success flag that is perfectly correlated for all parties, it is straightforward to generalize the ladder construction from \cref{sec:2kphotons} to the tripartite case. That is, we can arrange a scenario where Alice, Bob and Charlie perform $k$ parallel versions of the protocols described here, using $6k$ photons in $6k$ modes, and output the first successful event according to some pre-arranged ordering. By construction they will always be outputting data corresponding to the same successful event, unless all events fail. This achieves values of
\begin{align}
    S_{\textrm{Merm}} & = \frac{7}{2} - \frac{3}{2}\left(\frac{5}{9}\right)^k,\\
    S_{\textrm{Svet}} & = 5 - \left(\frac{5}{9}\right)^k, \\
    S_{\textrm{CWR}} & = 10.54 - 0.54\left(\frac{5}{9}\right)^k.
\end{align}

\begin{figure}
    \centering
    \includegraphics[trim=0 0 -50 0, width=0.65\linewidth]{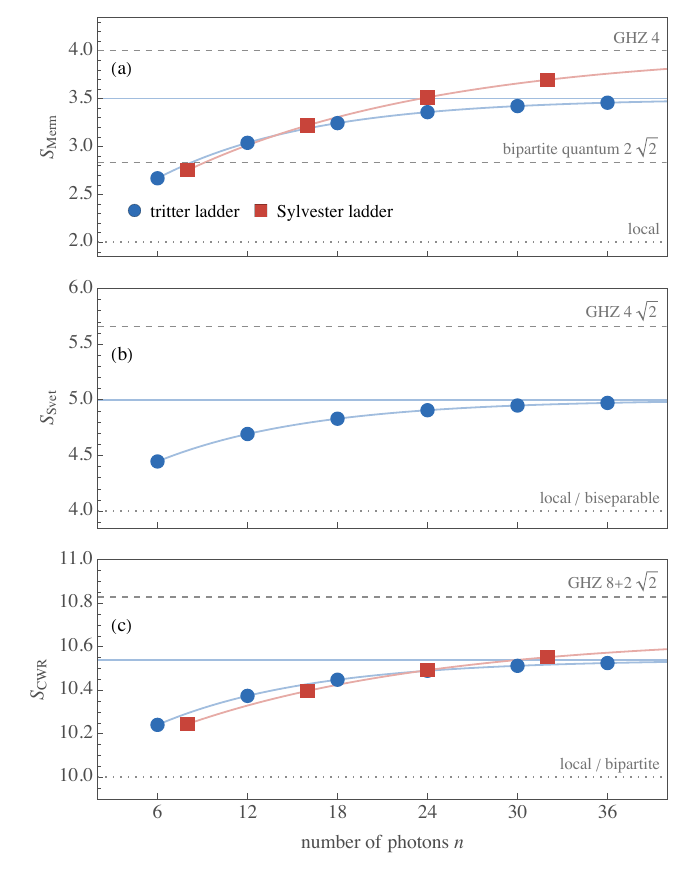}
    \caption{Violations of Mermin, Svetlichny and CWR inequalities as a function of the photon number. Points represent the tritter ladder (circles) from \cref{sec:tripartite} and the Sylvester ladder (squares) from \cref{app:sylvester}. 
    Light curves follow the points as functions of continuous $n$; horizontal solid lines mark the asymptotic values of the tritter ladders.}
    \label{fig:tripartitevsn}
\end{figure}

Note that the three values converge exponentially fast to $S_{\textrm{Merm}}=3.5$,  $S_{\textrm{Svet}}=5$ and $S_{\textrm{CWR}} = 10.54$, which are all violations of their corresponding bipartite bounds, though they all fall short of the corresponding maximum achieved by quantum mechanics. Interestingly, $S_{\textrm{Merm}}$ does not surpass the $2\sqrt{2}$ bound achievable by (quantum) bipartite resources for $k=1$, but it does so for $k=2$. In \cref{fig:tripartitevsn} we plot the best violations obtained as functions of the total photon number.

The value $7/2$ for Mermin's inequality is not a limitation of postselection-free
linear optics, but of the angle obstruction described above, which seems to be
specific to the tritter. In Appendix~\ref{app:sylvester} we show that, if we replace
the two tritters by two four-mode Sylvester interferometers (for a total of 8 photons in 8 modes)
and give one of the parties two pairs of modes, every successful event, now
occurring with probability $3/8$, carries a GHZ state on which the same phase
shifters act consistently, irrespective of the encoding. This gives
$S_{\rm Merm}=4$ in the successful sector, and the ladder construction then gives
\begin{equation} \label{eq:mermmax}
  S_{\rm Merm} = 4 - 2\left(\frac{5}{8}\right)^{k},
\end{equation}
which converges exponentially fast to the algebraic maximum. The same construction does not improve on the Svetlichny value, but it does improve the CWR value, giving
\begin{equation} \label{eq:cwrsylv}
  S_{\rm CWR} = 10.65 - 0.65\left(\frac{5}{8}\right)^{k},
\end{equation}
which converges to a value slightly above that of the tritter scheme.

\section{Conclusions}

In this work, we showed how we can use pure passive linear optics with no postselection to violate the CHSH inequality beyond the value $1+\sqrt{2}$, which we proved optimal when the entangled state is prepared by an interferometer acting on a two-photon Fock state. Our construction yields violations arbitrarily close to the Tsirelson bound, at the cost of additional photons. We also showed how a similar idea, but using a larger interferometer, can demonstrate genuine tripartite nonlocality via violations of the CWR inequality beyond its bipartite bound{, and similar violations for Mermin and Svetlichny.}

The main takeaway from our work is that Fock-state linear optics is not limited to any CHSH value below Tsirelson's bound, even in the absence of postselection: the limit of $1+\sqrt2$ holds for the two-photon case, but can be surpassed at the cost of using more photons.  This is in contrast with its \emph{computational} capabilities, where linear optics is known to be universal when supplemented with postselection or measurement adaptivity, but without them reduces to the much more restricted model of BosonSampling.

Finally, we comment on experimental requirements. A major limitation of our constructions is photon loss, since they all rely on each party learning, from their own photon count, whether a run succeeded. This, in turn, exploits the fact that the success flags are perfectly correlated across all parties. A lost photon might break that correlation, in the same way that it breaks the benign-postselection argument discussed in \cref{sec:intro}. Because we keep all the data, however, losses translate directly into a detection-efficiency threshold. For the two-photon scheme of \cref{sec:background}, for example, this threshold is known: for a uniform detection efficiency $\eta$, the violation survives whenever $\eta > 4/(3+\sqrt{2}) \approx 0.906$
\cite{Sciarrino2011,gebhart2022,zukowski_irrelevance_1999}. 

That said, loss is less damaging for the ladder construction of \cref{sec:2kphotons} than one might expect. Since the parties only report the first copy that each of them flags as successful, losses in later copies are irrelevant, and a simple calculation for uniform, independent photon loss shows that the ladder violates the CHSH inequality for $\eta > 4/(3+\sqrt{2})$ for every $k$, with a violation that still increases with $k$, up to
$2\sqrt{2}\,\eta/(4-3\eta)$.\footnote{With each photon detected independently
with probability $\eta$, a copy of the scheme of \cref{fig:ladder} gives one of three
types of event: (i) both parties detect one photon, a genuine success, with
probability $s=\eta^2/2$, (ii) both parties register a failure and move on to the
next copy with probability $f=(1-\eta)^2+\eta^2/2$, and (iii) one party detects exactly
one photon while the other detects none with probability $d=2\eta(1-\eta)$.
In the last case the detected photon has either lost its partner or belongs
to a pair that went to a single party, so its party's output is uniformly
random and independent of the other's, and these runs contribute
nothing to $S_{\rm CHSH}$. Hence
$S_{\rm CHSH}=2\sqrt{2}\,s\,(1-f^k)/(1-f)+2f^k$, which, since $1-f=s+d$,
exceeds 2 if and only if $\sqrt{2}\,s>s+d$, i.e., $\eta>4/(3+\sqrt{2})$.}

Another important issue is partial photon distinguishability, which we expect to be particularly harmful to the tripartite constructions, since their success flags rely on suppression laws that require all photons to be indistinguishable. We leave a full analysis of the effects of both loss and partial distinguishability in the tripartite constructions for future work. 

Our constructions also require photon-number-resolving detection, since
the success flags are the parties' local photon counts (in the tritter
scheme, for instance, a party must distinguish between one and four photons). In some
two-photon setups this requirement can be partly relaxed \cite{zukowski_irrelevance_1999},
but implementing the flags of our larger schemes requires number resolution,
or pseudo-number resolution obtained by distributing each mode over several
threshold detectors.

On the positive side, all the ingredients of our schemes have been
demonstrated in recent small-scale experiments. The two-photon scheme of
\cref{sec:background} was implemented with number-resolving detection in
Ref.~\cite{kun2026}, and the three-photon construction of \cref{apx:odd} would
surpass the two-photon bound of \cref{sec:newbound} with a single additional photon.
Three-photon interference in tritters has been used to generate
GHZ- and W-type states \cite{kumar_experimental_2023}, the suppression law of Sylvester
interferometers underlying \cref{app:sylvester} has been observed in integrated four-
and eight-mode devices \cite{viggianiello_experimental_2018}, and pseudo-number-resolving
detection has been used in multiphoton experiments \cite{rodari_experimental_2026}. We hope
our results motivate demonstrations of postselection-free nonlocality on
these platforms.

\begin{acknowledgments}
We acknowledge the use of Claude Opus 5.5 and Fable 5.1 (Anthropic) and Gemini Carbon (Google) as an aid in all the numerical optimization work, and in polishing the text and figures of this manuscript. The final writing, the mathematical results, their derivations, and their correctness were independently verified by the authors, who take full responsibility for the technical accuracy and integrity. EFG acknowledges support from the National Council for
Scientific and Technological Development – CNPq (Brazil) under grant 308292/2025-1. DJB acknowledges support from National Council for
Scientific and Technological Development – CNPq (Brazil) under grant 303446/2025-0 and from Carlos Chagas Filho Foundation for Research Support in the State of Rio de Janeiro (FAPERJ) for its financial support under grant E-26/204.533/2025 (315946).
\end{acknowledgments}

\appendix
\crefalias{section}{appendix}
\crefalias{subsection}{subappendix}

\section{CHSH with odd photon numbers} \label{apx:odd}

In this Appendix we show an alternative construction with odd numbers of photons that modestly improves on the $2k$-photon strategy described in \cref{sec:2kphotons}. Besides the general improvement, this also works for the $n=3$ case specifically, which might be interesting for current small-scale experiments \cite{kumar_experimental_2023,rodari_experimental_2026}.

\subsection{3 photons in 4 modes}

\begin{figure}
    \centering
    \includegraphics[width=0.8\linewidth]{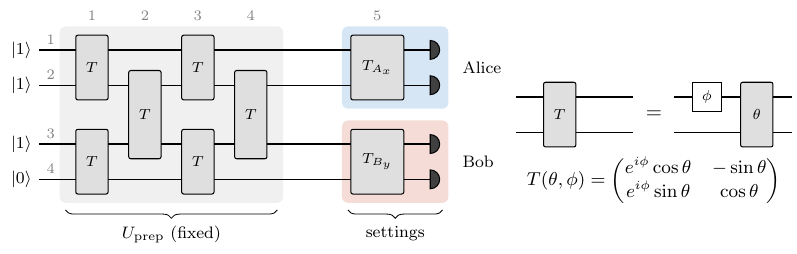}
    \caption{Interferometer for CHSH violation with $n=3$ photons.}
    \label{fig:chshwith3}
\end{figure}

We begin with a construction that uses three photons in four modes, and which reaches
\begin{equation} \label{eq:bestn3}
  S_{\textrm{CHSH}} \;=\; {2.5164}\;>\;1+\sqrt2 \approx 2.4142.
\end{equation}
The interferometer is a universal rectangular circuit in the Clements arrangement \cite{clements2016,pentangelo_highfidelity_2024} on four modes: beam splitters between neighboring modes only, in alternating columns. The initial state consists of three single photons on modes $1,2,3$ and vacuum on mode $4$. Layers $1$--$4$ implement a general four-mode unitary $U_{\rm prep}$, and the last layer consists of the measurement settings on Alice's and Bob's sides, as in \cref{fig:chshwith3}.

\begin{table}[h] 
\centering\small
\begin{tabular}[t]{cccc}
\toprule
\multicolumn{4}{c}{Preparation ($U_{\rm prep}$)}\\
\cmidrule(lr){1-4}
layer & modes & $\theta$ & $\phi$\\
\midrule
1 & $(1,2)$ & 3.5381 & 3.9177\\
1 & $(3,4)$ & 1.9220 & 1.2071\\
2 & $(2,3)$ & 5.5289 & 5.4393\\
3 & $(1,2)$ & 2.1726 & 0.8122\\
3 & $(3,4)$ & 1.2234 & 4.1406\\
4 & $(2,3)$ & 0.5766 & 2.3599\\
\bottomrule
\end{tabular}
\qquad
\begin{tabular}[t]{cccc}
\toprule
\multicolumn{4}{c}{Settings}\\
\cmidrule(lr){1-4}
setting & modes & $\theta$ & $\phi$\\
\midrule
$A_0$ & $(1,2)$ & 0.4904 & 2.4856\\
$A_1$ & $(1,2)$ & 1.0427 & 3.1587\\
$B_0$ & $(3,4)$ & 0.9339 & 4.5281\\
$B_1$ & $(3,4)$ & 0.5945 & 0.0536\\
\bottomrule
\end{tabular}
\caption{Cell parameters of the interferometer of \cref{fig:chshwith3}, in the convention $T(\theta,\phi)$.}
\label{tab:threephoton}
\end{table}

Explicitly, with the convention that each two-mode element is of the type
$T(\theta,\phi)=\left(\begin{smallmatrix}e^{i\phi}\cos\theta & -\sin\theta\\ e^{i\phi}\sin\theta & \cos\theta\end{smallmatrix}\right)$, the ten cell settings are given in \Cref{tab:threephoton}.
With these settings, we obtain
$$S_{\textrm{CHSH}}=2.5164.$$
This is the highest violation we could find with three photons in four modes. The sign assignment is important to achieve this value. Each party outputs $+1$ on every count pattern except those listed:
\begin{center}\small
\begin{tabular}{ll}
$A_0$: & $-1$ on $(0,0),(1,0),(1,2),(2,0),(2,1),(3,0)$\\
$A_1$: & $-1$ on $(0,1),(2,0)$\\
$B_0$:   & $-1$ on $(0,2),(1,0)$\\
$B_1$:   & $-1$ on $(0,0),(0,2),(0,3),(1,0),(1,1),(1,2),(2,1),(3,0)$\\
\end{tabular}
\end{center}
\noindent with patterns written $(n_1,n_2)$ for Alice and $(n_3,n_4)$ for Bob. 

The explicit unitary matrix, obtained by multiplying the six preparation cells above, is
{\small\setlength{\arraycolsep}{3pt}
\[
U_{\rm prep}=\left(\begin{array}{rrrr}
-0.22232-0.50641i & +0.21785-0.57275i & +0.06908+0.18149i & +0.53001\phantom{+0.00000i}\\
-0.38215-0.59988i & -0.13346+0.32853i & +0.09699+0.36624i & -0.46165+0.10817i\\
-0.18442-0.30394i & -0.34023+0.17402i & -0.37198-0.66413i & +0.23602+0.30403i\\
+0.14841+0.19951i & -0.58678-0.09176i & -0.02618+0.48850i & +0.23104+0.54106i
\end{array}\right).
\]}
The fourth column is never occupied, since mode 4 is initialized in the vacuum. 

\subsection{Beyond $n=3$}

The three-photon result can be combined with the $2k$-photon ladder of \cref{sec:2kphotons}. For odd $n=2k+3$, the best strategy we found is to run a $2k$-photon ladder identical to \cref{sec:2kphotons}, plus one final three-photon block on four final modes. Each party then reports the outcomes of its first successful pair and, when all pairs fail, they report the outcome of the final 3-photon block. This leads to a violation of $S_{\textrm{CHSH}}=(1-2^{-k})2\sqrt2+2^{-k}S_3$, where $S_3$ is the best value obtained in \cref{eq:bestn3}. At $n=5$ photons (in 8 modes) that is $S_{\textrm{CHSH}}=2.6724$, against $2.6213$ for the four-photon ladder.

\section{Measurement settings for the tripartite scheme} \label{apx:tripartite}

{
In this Appendix we list the measurement settings that achieve the values for the tripartite correlators described in \cref{sec:tripartite}. Each party receives two modes, one output of each tritter (cf.\ \cref{fig:tripartite}). To implement setting $x$, a party applies the transformation $T(\theta^{(x)},\phi^{(x)})$ in the notation of \cref{apx:odd}, i.e., a phase $\phi^{(x)}$ on the mode coming from the first tritter followed by a beam splitter, and then counts the photons $(n_1,n_2)$ at the two outputs. Writing $n=n_1+n_2$ for the total number of photons received, the party's output is a function of $x$, $n$ and $n_1$, given in \cref{tab:svet,tab:merm,tab:cwr}. The columns $n=1$ and $n=4$ correspond to successful runs, and the column ``failure'' to $n\in\{0,2,3,6\}$.

For $S_{\textrm{Svet}}$ and $S_{\textrm{Merm}}$, all beam splitters are balanced, $\theta^{(x)}=\pi/4$. For $n=1$ and $n=4$, the parity of $n_1$ then implements the equatorial qubit measurement described in \cref{sec:tripartite}, in the dual-rail and in the $\{\ket{31},\ket{13}\}$ encodings, respectively. The sign in front of the four-photon parity relabels the four-photon outcome. Since a party learns whether it holds one or four photons at detection, this is local post-processing, but it can differ between the two settings, which shifts the effective four-photon measurement angle by $\pi$ for that setting only. This freedom is needed for the Svetlichny value: without it, the best value we find in the successful sector is $9/2$ instead of $5$. On failed runs, Alice outputs $(-1)^x$ and Bob and Charlie output $+1$, which saturates the local bounds of both inequalities.

\begin{table}[h]
\centering\small
\setlength{\tabcolsep}{9pt}
\renewcommand{\arraystretch}{1.15}
\begin{tabular}{lcccccc}
\toprule
 & setting & $\theta$ & $\phi$ & $n=1$ & $n=4$ & failure \\
\midrule
Alice   & 0 & $\pi/4$ & $\pi/2$  & $(-1)^{n_1}$ & $(-1)^{n_1}$  & $+1$ \\
        & 1 & $\pi/4$ & $7\pi/6$ & $(-1)^{n_1}$ & $(-1)^{n_1}$  & $-1$ \\
\cmidrule{1-7}
Bob     & 0 & $\pi/4$ & $\pi/2$  & $(-1)^{n_1}$ & $(-1)^{n_1}$  & $+1$ \\
        & 1 & $\pi/4$ & $5\pi/6$ & $(-1)^{n_1}$ & $-(-1)^{n_1}$ & $+1$ \\
\cmidrule{1-7}
Charlie & 0 & $\pi/4$ & $\pi/6$  & $(-1)^{n_1}$ & $-(-1)^{n_1}$ & $+1$ \\
        & 1 & $\pi/4$ & $5\pi/6$ & $(-1)^{n_1}$ & $-(-1)^{n_1}$ & $+1$ \\
\bottomrule
\end{tabular}
\caption{Settings and outputs for the Svetlichny strategy.}
\label{tab:svet}
\end{table}

The settings of \Cref{tab:svet} give $S_{\textrm{Svet}}=5$ in the successful sector and $4$ in the failed sector, which occur with probabilities $4/9$ and $5/9$, respectively, leading to $S_{\textrm{Svet}}=40/9\approx4.44$.

\begin{table}[h]
\centering\small
\setlength{\tabcolsep}{9pt}
\renewcommand{\arraystretch}{1.15}
\begin{tabular}{lcccccc}
\toprule
 & setting & $\theta$ & $\phi$ & $n=1$ & $n=4$ & failure \\
\midrule
Alice   & 0 & $\pi/4$ & $\pi$    & $(-1)^{n_1}$ & $-(-1)^{n_1}$ & $+1$ \\
        & 1 & $\pi/4$ & $4\pi/3$ & $(-1)^{n_1}$ & $(-1)^{n_1}$  & $-1$ \\
\cmidrule{1-7}
Bob     & 0 & $\pi/4$ & $\pi$    & $(-1)^{n_1}$ & $-(-1)^{n_1}$ & $+1$ \\
        & 1 & $\pi/4$ & $4\pi/3$ & $(-1)^{n_1}$ & $(-1)^{n_1}$  & $+1$ \\
\cmidrule{1-7}
Charlie & 0 & $\pi/4$ & $0$      & $(-1)^{n_1}$ & $(-1)^{n_1}$  & $+1$ \\
        & 1 & $\pi/4$ & $2\pi/3$ & $(-1)^{n_1}$ & $(-1)^{n_1}$  & $+1$ \\
\bottomrule
\end{tabular}
\caption{Settings and outputs for the Mermin strategy.}
\label{tab:merm}
\end{table}

The settings of \Cref{tab:merm} give $S_{\textrm{Merm}}=7/2$ in the successful sector and $2$ in the failed sector, leading to $S_{\textrm{Merm}}=8/3\approx2.67$. 

The settings that maximize $S_{\textrm{CWR}}$ are qualitatively different (\Cref{tab:cwr}): all phases are $\phi^{(x)}=0$, and the beam-splitter angle $\theta^{(x)}$ varies instead. For $\theta=0$ there is no beam splitter, and the measurement is in the computational basis of both encodings, while for $\theta=\pi/4$ it is the $\sigma_x$ measurement of the Mermin and Svetlichny strategies, again in both encodings. Bob's settings 0 and 1 use unbalanced beam splitters. These implement measurements at angles $\pm\pi/4$ from the $Z$ axis in the dual-rail encoding, but do not act as qubit measurements on $\{\ket{31},\ket{13}\}$, and can cause the state to leak out of the encoding. Their four-photon outputs are therefore not a relabelled parity, and we specify them through the function $h(n_1)$, with $h=+1$ for $n_1\in\{0,1,3\}$ and $h=-1$ for $n_1\in\{2,4\}$. On failed runs every party outputs $+1$, except Charlie on setting 1, who outputs a uniformly random bit. Since his outcome is unbiased in successful runs, this ensures $\langle C_1\rangle = 0$. These settings give $S_{\textrm{CWR}} = 10.54$ in the successful sector and 10 in the failed sector, leading to $S_{\textrm{CWR}} \approx 10.24$.

\begin{table}[h]
\centering\small
\setlength{\tabcolsep}{9pt}
\renewcommand{\arraystretch}{1.15}
\begin{tabular}{lcccccc}
\toprule
 & setting & $\theta$ & $\phi$ & $n=1$ & $n=4$ & failure \\
\midrule
Alice   & 0 & $0$       & $0$ & $(-1)^{n_1}$  & $(-1)^{(n_1+1)/2}$ & $+1$ \\
        & 1 & $\pi/4$   & $0$ & $(-1)^{n_1}$  & $(-1)^{n_1}$  & $+1$ \\
\cmidrule{1-7}
Bob     & 0 & $-3\pi/8$ & $0$ & $-(-1)^{n_1}$ & $h(n_1)$      & $+1$ \\
        & 1 & $-\pi/8$  & $0$ & $(-1)^{n_1}$  & $-h(n_1)$     & $+1$ \\
        & 2 & $0$       & $0$ & $(-1)^{n_1}$  & $(-1)^{(n_1+1)/2}$ & $+1$ \\
\cmidrule{1-7}
Charlie & 0 & $0$       & $0$ & $(-1)^{n_1}$  & $(-1)^{(n_1+1)/2}$ & $+1$ \\
        & 1 & $\pi/4$   & $0$ & $(-1)^{n_1}$  & $(-1)^{n_1}$  & random \\
\bottomrule
\end{tabular}
\caption{Settings and outputs for the CWR strategy.}
\label{tab:cwr}
\end{table}

In the parallel construction with $k$ copies, each party uses the same settings in every copy, for all three inequalities.}

\section{Maximal Mermin value with Sylvester interferometers}
\label{app:sylvester}

In this Appendix we describe the construction mentioned at the end of
Section~\ref{sec:tripartite}. We begin by revisiting the obstruction in the
tritter scheme.

Consider a pair of modes in the span of $|a,b\rangle$ and $|b,a\rangle$, with
$a\neq b$, and define its \emph{charge} $q=a-b$. The Mermin and Svetlichny measurements of
Appendix~\ref{apx:tripartite}, i.e., the
transformation $T(\pi/4,\phi)$ followed by recording $(-1)^{n_1}$, are then projective
measurements of
\begin{equation}
  \cos(q\phi)\,\sigma_x + \sin(q\phi)\,\sigma_y
  \label{eq:charge}
\end{equation}
in the basis $\ket{0}=\ket{a,b}$, $\ket{1}=\ket{b,a}$. To see this, note
first that for $\phi=0$ the beam splitter maps $\ket{a,b}$ and $\ket{b,a}$ to
states related by $b_1^\dagger\mapsto -b_1^\dagger$, i.e., by the action of
$(-1)^{\hat n_1}$, so the image of $\ket{a,b}\pm\ket{b,a}$ contains only even
(odd) $n_1$. The phase then multiplies $\ket{a,b}$ by $e^{ia\phi}$ and
$\ket{b,a}$ by $e^{ib\phi}$, which rotates the measurement by $q\phi$. The
dual-rail and $\{\ket{31},\ket{13}\}$ encodings of Section~\ref{sec:tripartite}
correspond to $|q|=1$ and $|q|=2$, respectively, and Eq.~\eqref{eq:charge}
reproduces the angle doubling. In each $\ket{\mathrm{GHZ}_i}$ event of the tritter scheme,
the occupations of one tritter's outputs differ between the two branches by
$(2,-1,-1)$ up to permutation. Three integers that sum to zero cannot all be odd,
so in every event some party holds a qubit of even charge, and each party plays
this role in one of the three events. This is what prevents the tritter scheme
from reaching $S_{\rm Merm}=4$.

\begin{figure}
    \centering
    \includegraphics[width=0.5\linewidth]{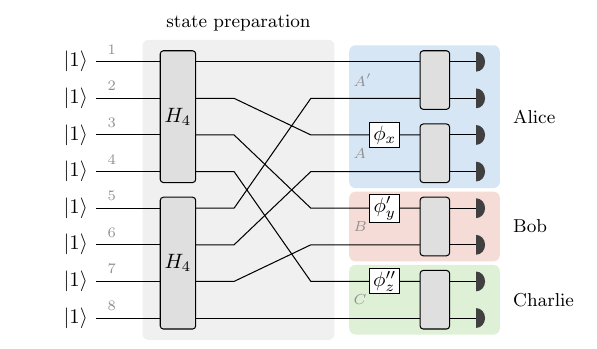}
    \caption{Circuit for violation of the Mermin and CWR inequalities with 8 photons, 8 modes, and pure passive linear optics.}
    \label{fig:sylvester}
\end{figure}

Four odd integers, however, can sum to zero. Consider the four-mode Sylvester
interferometer $H_4=H\otimes H$, with $H$ the balanced beam splitter, whose matrix
elements are $(H_4)_{mj}=(-1)^{m\cdot j}/2$ when modes are labelled by two-bit
strings $m,j\in\{00,01,10,11\}$. We use two copies of it, each with one
photon per input mode, for a total of 8 photons in 8 modes, and distribute the
outputs as in \cref{fig:sylvester}. Each party receives pairs consisting of the
outputs with the same label $m$ in the two blocks: Alice receives the pairs
$A$ ($m=01$) and $A'$ ($m=00$), Bob receives $B$ ($m=10$), and Charlie receives
$C$ ($m=11$).

With input $|1111\rangle$, the Sylvester suppression law
\cite{crespi2015,viggianiello_experimental_2018,dittel_totally_2018} states that an output pattern
$(n_{00},n_{01},n_{10},n_{11})$ has nonzero probability only if $n_{01}+n_{11}$ and
$n_{10}+n_{11}$ are both even. The populated patterns are $|1111\rangle$, with
probability $1/4$; the four patterns with all photons in one mode, with probability
$3/32$ each; and the six patterns with two photons in each of two modes, with
probability $1/16$ each. In each block, $n_{01}$, $n_{10}$ and $n_{11}$ therefore
have the same parity, and they are all odd only for $|1111\rangle$. As a
consequence, the photon numbers $N_A$, $N_B$, $N_C$ of the three pairs have the
same parity in every run. This is the failure flag of this strategy: each party
declares success when its pair (pair $A$, for Alice) holds an odd number of
photons, which happens when exactly one block outputs $|1111\rangle$, with
probability $p_s = 2\cdot\frac14\cdot\frac34 = 3/8$. Note that Alice must not
interfere her two pairs with each other, since $N_A+N_{A'}$ is even in every run.

In a successful run, one block outputs $|1111\rangle$ and the other an even pattern
$e=(e_{00},e_{01},e_{10},e_{11})$. Since the two blocks are identical, both
possibilities have the same amplitude, and the state conditioned on $e$ is
\begin{equation}
  \ket{\mathrm{GHZ}_e}= \frac{1}{\sqrt2}\left(\bigotimes_X |1,e_X\rangle_X
  + \bigotimes_X |e_X,1\rangle_X\right),
  \label{eq:ghz-e}
\end{equation}
a four-qubit GHZ state of which Alice holds two qubits. Crucially, every pair now
has odd charge, $q_X = 1-e_X = 2-N_X \in\{1,-1,-3\}$.

Each party uses phases $\phi^{(0)}=0$ and $\phi^{(1)}=\pi/2$ on its pair (Alice on
pair $A$, while pair $A'$ always uses phase $0$). By Eq.~\eqref{eq:charge}, a qubit
of odd charge is then measured in $\sigma_x$ for setting $0$, and in
$\sin(q\pi/2)\,\sigma_y = (-1)^{(N-1)/2}\sigma_y$ for setting $1$, where the sign is
known to each party from its own photon count. On success, a party with setting
$x$ outputs
\begin{equation}
  o^{(x)} = f^{(x)}(N)\prod_X (-1)^{n_1^{(X)}}, \qquad
  f^{(0)}=1, \quad f^{(1)}(N) = (-1)^{(N-1)/2},
\end{equation}
where the product runs over the party's pairs. On failure, Alice outputs $(-1)^x$
and Bob and Charlie output $+1$, which saturates the local bound $S_{\rm Merm}=2$.
For the state of Eq.~\eqref{eq:ghz-e},
$\langle\bigotimes_X(\cos\theta_X\sigma_x+\sin\theta_X\sigma_y)\rangle
=\cos\left(\sum_X\theta_X\right)$, which gives $\langle A_0B_0C_0\rangle=1$ and
$\langle A_0B_1C_1\rangle=\langle A_1B_1C_0\rangle=\langle A_1B_0C_1\rangle=-1$,
i.e., $S_{\rm Merm}=4$ in every successful sector. Keeping all data, a single copy
gives $S_{\rm Merm}=\frac38\cdot 4+\frac58\cdot 2 = 11/4$, and the ladder
construction gives \cref{eq:mermmax}. As in the tritter case, one copy does not exceed
$2\sqrt2$, but two copies (16 photons) do, and three copies (24 photons) already
exceed $7/2$. 

No similar construction was found to improve on the Svetlichny value. The standard GHZ
strategy for Svetlichny's inequality requires settings at angles $\pm\pi/4$, and a
phase $\pm\pi/4$ rotates a qubit of charge $q$ by $\pm q\pi/4$. For
$q\equiv 1 \pmod 4$ this reproduces the two angles up to signs, but for
$q\equiv 3 \pmod 4$ it exchanges them, and relabelling outcomes cannot undo the
exchange. Here $q\equiv1$ for $N\in\{1,5\}$ and $q\equiv3$ for $N=3$, so every pair
takes both kinds of value. Optimizing numerically over all phases and output signs,
the best value we found in the successful sector is $S_{\rm Svet}=4.344$, below the
value $5$ of the tritter scheme.

The Sylvester construction does, however, improve on the CWR value relative to the tritter scheme. On the settings for $S_\textrm{same}$ ($A_0$, $B_2$ and $C_0$) there is no beam splitter, and on success a party outputs $+1$ if its pair (pair $A$ for Alice) holds exactly one photon in the mode coming from the first block, and $-1$ otherwise. On setting 1, Alice and Charlie use balanced beam splitters with no phase on all their pairs and output $\prod_X (-1)^{n_1^{(X)}+(N_X-1)/2}$, as in Eq.~(C3). On settings 0 and 1, Bob uses $T(-0.587\pi,0)$ and $T(0.587\pi,0)$, respectively, and outputs $-\mathrm{sgn}(n_1-n_2)$ if $N_B=1$ and $\mathrm{sgn}(n_1-n_2)$ otherwise. On failure, every party outputs $+1$, except Charlie on setting 1, who outputs a uniformly random bit. Since every pair has odd charge, Bob's unbalanced beam splitters now violate CHSH in every successful event, which gives $S_{\textrm{CWR}} \approx 10.65$ in the successful sector, against $10.54$ for the tritter. Keeping all data, a single copy gives $S_{\textrm{CWR}} \approx 10.24$, and the ladder construction gives Eq.~\eqref{eq:cwrsylv}.

\bibliography{refs}
\end{document}